# Formation of a Buffer Layer for Graphene on C-face SiC{0001}

Guowei He, N. Srivastava, and R. M. Feenstra
Dept. Physics, Carnegie Mellon University, Pittsburgh, PA 15213

**Abstract**

Graphene films prepared by heating the $SiC(000\bar{1})$ surface (the *C-face* of the {0001} surfaces) in a Si-rich environment are studied using low-energy electron diffraction (LEED) and low-energy electron microscopy (LEEM). Upon graphitization, an interface with $\sqrt{43} \times \sqrt{43}$-R±7.6° symmetry is observed by *in situ* LEED. After oxidation, the interface displays $\sqrt{3} \times \sqrt{3}$-R30° symmetry. Electron reflectivity measurements indicate that these interface structures arise from a graphene-like "buffer layer" that forms between the graphene and the SiC, similar to that observed on Si-face SiC. From a dynamical LEED structure calculation for the oxidized C-face surface, it is found to consist of a graphene layer sitting on top of a silicate ($Si_2O_3$) layer, with the silicate layer having the well-known structure as previously studied on bare $SiC(000\bar{1})$ surfaces. Based on this result, the structure of the interface prior to oxidation is discussed.

## I. Introduction

Graphene, a single sheet of *sp$^2$*-bonded carbon arranged in a honeycomb lattice, has potential for novel electronic devices due to its unusual electronic properties.[1,2,3] Formation of graphene on SiC has been intensively studied for the past several years, since graphene formed in that way can have large areas suitable for device and circuit fabrication.[3] There are two inequivalent faces of SiC{0001}: the (0001) face which is known as the Si-face, and the $(000\bar{1})$ face known as the C-face. On both of these surfaces, by heating to temperatures at about 1200 °C, Si atoms preferentially sublimate from the surface, leaving behind excess C atoms that self-assemble into graphene. On the Si-face, a number of groups have succeeded in forming single layer of graphene, with good reproducibility between groups.[3,4,5,6] In contrast, for the C-face, a number of studies reveal the formation of islands of graphene instead of a uniform single layer.[7,8,9]

For graphene on SiC, it has been demonstrated that new graphene layers are formed not *on top* of existing ones, but rather, they form at the *interface* between existing graphene layers and the underlying substrate.[10] Hence, the starting surface of SiC and the later interface structure between the graphene film and the SiC substrate play a crucial role for subsequent graphene formation. To date the graphene/SiC interface is quite well understood for the Si-face: the interface consists of a C-rich layer having $6\sqrt{3} \times 6\sqrt{3}$-R30° symmetry (denoted $6\sqrt{3}$ for short), which is covalently bonded to the underlying SiC substrate.[10,11] This interface on the Si-face acts as an electronic "buffer" layer between graphene films and SiC substrate and provides a template for subsequent graphene formation.[12] By the term *buffer layer* here, we mean a layer that has nearly the same structure as graphene, but is covalently bonded to the underlying material and therefore has different electronic structure than graphene.[12] This Si-face buffer layer is observed by several groups from samples prepared under various preparation conditions.[3]

For the C-face, in contrast, the formation of interface structures and subsequent graphene films strongly depends on preparation conditions. Most studies reveal 3 × 3 and/or 2 × 2 interface structures, which are believed to *not* act as an electronic buffer layer nor provide a template for subsequent graphene formation.[10,13] In addition, oxidation of the surface before graphitization brings in another difference between the Si-face and the C-face: as we have demonstrated previously, the C-face is much more susceptible to oxidation, yielding a silicate layer on the surface which inhibits the formation of graphene.[9] To achieve better quality of graphene films on the C-face, more detailed studies of the interface structures and the relationship between these structures and graphene preparation conditions are needed.

In our prior work, we prepared graphene on the C-face of SiC in Si-rich environments, utilizing either disilane at a pressure of ~$10^{-4}$ Torr, or cryogenically purified neon at 1-atm pressure. We find that when graphene is prepared in these preparation conditions, a new interface structure with $\sqrt{43} \times \sqrt{43}$-R±7.6° symmetry is found.[14,15,16] After subsequent oxidation of the surface by mild heating in the presence of oxygen, the structure transforms to one with $\sqrt{3} \times \sqrt{3}$-R30° symmetry. We have previously argued that both the $\sqrt{43} \times \sqrt{43}$-R±7.6° and $\sqrt{3} \times \sqrt{3}$-R30° structures are indicative of a graphene-like buffer layer that terminates the SiC crystal.[15,16] That is, with additional graphene formation on the surface, this buffer layer is present at the interface between the graphene and the SiC, just as occurs for the Si-face surface.

In this work, we discuss the formation and structure of the C-face buffer layer, providing new results to illustrate its characteristics. First, we summarize prior results for low-energy electron diffraction (LEED) patterns and low-energy electron reflectivity (LEER) spectra of the buffer layer. Both types of data were presented in our prior work, but a complete understanding of the LEER spectra in particular was not available at that time. We subsequently developed a first-principle method for computing such spectra,[17,18] and based on that we can now provide a more rigorous interpretation of those spectra. Following that presentation, we then describe a quantitative LEED structure determination for the observed $\sqrt{3} \times \sqrt{3}$-R30° surface (i.e. the C-face buffer following oxidation), revealing that its structure consists of a graphene layer on top of a $Si_2O_3$ silicate layer. Based on that result, we discuss the structure of the C-face buffer layer *prior to* the oxidation.

This paper is organized as follows. In Sec. II, we present details of our experimental and computational methods. Section III(A) describes our results from experimental LEED and LEER observations, including presentation of structural models and definition of the notation we use to refer to specific layers of the structures. Section III(B) gives results of our theoretical calculations of the LEED intensity vs. voltage (*I-V*) characteristics for the $\sqrt{3} \times \sqrt{3}$-R30° structure (the C-face buffer following oxidation) and provides a discussion of the possible structure of the $\sqrt{43} \times \sqrt{43}$-R±7.6° surface (the C-face buffer prior to oxidation). In Section IV we discuss our results, and in particular we compare them with recent data from de Heer and co-workers dealing with graphitized C-face SiC.[19] Those workers have obtained exceptional electronic transport properties for their surfaces, and that work provided important motivation for our present study. We describe how the graphitized C-face SiC surface produced in our work may or may not be the same as those produced by de Heer and co-workers.[19] Finally, our results are summarized in Section V.

**II. Experimental and theoretical methods**

Experiments were performed on nominally on-axis, *n*-type 6H-SiC or semi-insulating 4H-SiC wafers purchased form Cree Corp., with no apparent differences between results for the two types of wafers. The wafers were cut into $1 \times 1$ cm$^2$ samples. To remove polishing damage, the samples were heated in either 1 atm of hydrogen at 1600 °C for 3 min or $5 \times 10^{-5}$ Torr of disilane at 850 °C for 5 min. In the same chamber, graphene was formed by heating in $5 \times 10^{-5}$ Torr of disilane. Characterization by low-energy electron diffraction (LEED) was performed *in situ* in a connected ultra-high vacuum (UHV) chamber.

For quantitative LEED analysis, diffraction spot intensities were measured at different energies in the range of 100 – 300 eV. For the SiC surface of specific termination, a single domain with only one orientation would give rise to a threefold symmetric LEED pattern in which the (10) and (01) spots have different intensity spectra. Since a six-fold symmetric LEED patterns are indeed observed, both possible domains with different orientations, i.e. rotated by 60° with respect to each other, are present on the surface. Spot intensities from two rotational domains were averaged and the resulting *I*(*E*) spectra were compared to theoretical LEED calculations in order to retrieve details about atomic arrangement of the interface structure. The theoretical *I*(*E*) is calculated by full dynamical LEED calculation and optimization was carried out by tensor LEED, using calculation package from Blum et al.[20] The Pendry R-factor, $R_p$,[21] was used for comparison between experimental and calculated *I*(*E*) spectra.

**III. Results**

**A. Structural models, LEED patterns, LEER spectra**

Figure 1 shows structural models for the two surfaces that are the topic of this paper – a graphene-like buffer layer on C-face SiC, and the same buffer layer on a surface which has been oxidized. In both figures, the buffer layer is the topmost layer of the surfaces, with a carbon atom density and arrangement similar to that of graphene. The term "buffer layer", which we denote as "B", is used in Fig. 1(a) to refer to this graphene-like layer since it bonds to the underlying SiC structure. Actually, the precise interface structure between the graphene and the SiC is not known, as indicated by the box with question marks in Fig. 1(a). However, what *is* known is that the nature of the bonding between the buffer layer and the underlying SiC changes as a result of oxidation of the surface. As shown in Fig. 1(b), we find that *after* oxidation the SiC is terminated by a Si$_2$O$_3$ silicate and the buffer layer above that silicate is only weakly bonded to it. Hence, the buffer layer is *decoupled* from the underlying structure (analogous to what occurs on graphitized Si-face SiC[22,23,24,25]), and it forms a regular graphene layer which we refer to as "$G_0$" (with the subscript "0" referring to the fact that it originates from the buffer layer).

The model shown in Fig. 1(b) is actually the result of the detailed LEED *I-V* analysis of the following Section, but we introduce it here in advance of that analysis in order to provide some definiteness to the structures that we discuss. Figures 2(a) and 2(b) show LEED patterns acquired from the two surfaces corresponding to Figs. 1(a) and 1(b), respectively. The pattern of Fig. 2(a) was obtained from a surface *in situ* immediately after graphene preparation, which is done by heating the sample in $5 \times 10^{-5}$ Torr of disilane at 1250 °C for 5 min. Weak graphene streaks and a complex arrangement of spots are observed. As illustrated in our prior work,[15] the complex pattern can be indexed using a supercell with edges extending along (6,1) and (-1,7) of the SiC 1 × 1 cells. Using a compact notation we denote this structure as $\sqrt{43} \times \sqrt{43}$-R±7.6° (denoted by $\sqrt{43}$ for short). After this *in situ* study, the sample was exposed to air during transfer between preparation and characterization chambers, and after introduction into the LEEM chamber it was outgassed at about 1000 °C for several minutes. This procedure caused the $\sqrt{43}$ pattern to disappear and a $\sqrt{3} \times \sqrt{3}$-R30° pattern to appear, as shown in Fig. 2(b). The same $\sqrt{3} \times \sqrt{3}$-R30° pattern was found on samples that were exposed to $1 \times 10^{-5}$ Torr pure oxygen (rather than air) while heating to 1000 °C. So, the $\sqrt{3} \times \sqrt{3}$-R30° pattern is an indication of oxidation of the surface, as confirmed by the calculation in the follow section.

The LEED patterns of Fig. 2 provide a means of characterizing the unoxidized and oxidized buffer layer structures, although the LEED data suffers from the fact that the surfaces under study are not completely homogeneous (i.e. uniform from point to point over the surface area). The patterns of Fig. 2 were acquired with a conventional wide-area LEED apparatus, having an electron beam diameter of about 0.5 mm. Data of the sort shown in Fig. 2 was found to be fairly reproducible, i.e. displaying all the same features, for beam positions confined to the center 5x5 mm$^2$ of our samples. However, from studies with a low-energy electron microscopy (LEEM), we know that for any given 0.1x0.1 mm$^2$ area of the sample within this center region the graphene coverage on the surface varies. For an unoxidized sample, we observe the bare buffer layer (B) together with areas of buffer layer plus graphene (B+G) and occasional buffer layer plus more graphene layers (B+2G or B+4G). Similarly, for an oxidized sample, we observe areas of decoupled buffer layer which corresponds to a single graphene layer ($G_0$), together with areas of graphene on top of that ($G_0$+G) or areas with additional graphene layers. Most importantly, using the LEEM we have performed spatially resolved diffraction (μ-LEED) at many individual μm-sized locations over the center region of the samples. For samples displaying LEED patterns such as those Fig. 2, i.e. with well-developed $\sqrt{43}$ spots prior to oxidation, we find that *all* such locations display distinct graphene diffraction spots, arising either from the coupled or decoupled buffer layer or from graphene layer(s) on top of that. In addition to graphene diffraction spots, some locations also display the $\sqrt{3} \times \sqrt{3}$-R30° pattern as seen in Fig. 3(c) of our previous work.[15] This coexistence of both graphene and $\sqrt{3}$ spots again confirms that a well-ordered oxidation layer forms underneath the decoupled buffer layer.

LEER spectra measured with the LEEM provide a useful means of further characterizing the various layers on the sample surface. Figure 3 shows examples of such spectra, acquired from both unoxidized and oxidized samples.[16] These spectra of Fig. 3 can be easily interpreted if we bear in mind the recent interpretation that the *minima* in the spectra arise from electronic states localized between the graphene layers or between the bottommost layer and the substrate.[17,18] For *n* graphene layers there are *n*-1 spaces between them and, hence, *n*-1 interlayer states. An additional state forms between the bottommost graphene layer and the substrate so long as the space between those is sufficiently large. Coupling (in a tight-binding sense) between all the interlayer states then produces a set of coupled states, and reflectivity minima are observed at the energies of these coupled states.

For example, the LEER spectrum for the buffer (B) in Fig. 3(a) does not have any distinct minimum, since the buffer is relatively strongly bonded to the substrate and hence no interlayer state forms. For a layer of graphene on the buffer (B+G), a single state forms in the space between the buffer and the graphene and hence a single reflectivity minimum (at ~2.1 eV) results. Similarly, two minima form for B+2G and three minima for B+3G, with these sets of minima all approximately centered around 2 eV.

Turning to the oxidized surface, Fig. 3(b), the buffer layer now decouples from the substrate (forming a *decoupled buffer layer*, $G_0$) so that an interlayer state forms, with energy ~5.3 eV. The fact that this energy is higher than the 2.1 eV for the state between graphene layers indicates that the separation between the decoupled buffer and the substrate is *smaller* than that between two graphene layers (which is not surprising, since the graphene-graphene separation is likely close to a *maximum* interlayer separation considering the weak van der Waals bond between graphene layers).[17,18] For a graphene layer on the decoupled buffer ($G_0$+G), there are interlayer states at about 2.1 and 5.3 eV, and these do not have large coupling (due to their relatively large energy difference) so that reflectivity minima are observed at nearly the same energies.

The upper two spectra in Fig. 3(b) are essentially the same as the B+G and B+2G spectra of Fig. 3(a) and they are labeled as such. For the B+G spectrum of Fig. 3(b), we always find some evidence of that (along with the $G_0$+G spectra) on our oxidized surfaces, and we attribute the presence of the former simply to incomplete oxidation of the surface. For the case of the B+2G in Fig. 3(b), we cannot definitively distinguish that from a $G_0$+2G situation in which the bottommost interlayer state is not visible, but in any case for such spectra with two (or more) reflectivity minima centered around 2 eV we never observe any evidence of a higher reflectivity minimum near 5.3 eV. Interpreting such spectra as indeed arising from B+2G, it appears that oxidation of the SiC beneath multilayer graphene is more difficult than between single-layer graphene, a point that we return to in Section IV.

**B. LEED *I-V* analysis**

A primary goal of the present work is to learn about the structure of the graphene buffer layer on the C-face, as characterized by the LEED pattern of Fig. 2(a). However, that LEED pattern is very complex, being too complicated to permit dynamical LEED *I-V* analysis. For this reason, we focus on the pattern after oxidation of the sample, as shown in Fig. 2(b). The pattern is now relatively simple, with distinct $\sqrt{3} \times \sqrt{3}$-R30° spots. We have measured the *I-V* characteristics of those spots, with our experimental results shown by the solid lines in Fig. 4.

For comparison with the $\sqrt{3} \times \sqrt{3}$-R30° pattern of the oxidized (decoupled) buffer layer, we show in Fig. 5(a) a LEED pattern with the same symmetry obtained from a bare oxidized C-face SiC sample. This sample was *not* graphitized; rather, it was prepared by annealing the surface in a 1 atm argon environment with residual oxygen present. This pattern does not display any graphene spots, but it shows clear SiC 1 × 1 spots and $\sqrt{3} \times \sqrt{3}$-R30° spots that are indicative of oxidation.[9,26] LEED intensity vs. energy spectra for the various spots of this pattern are shown by the solid lines in Figs. 5(b) – 5(f). Also shown in those panels are the results of dynamical LEED calculations, which were carried out using a model that consists of one layer of silicate ($Si_2O_3$) and six layers of SiC bilayer. The geometry parameters of the $Si_2O_3$ layer are the same as that used by Starke *et al*.[26] We note that their analysis was done for various different surface terminations of the 6H SiC surfaces, i.e. S1, S2, and S2 referring to 1, 2, or 3 SiC bilayers, respectively, stacked in a cubic arrangement before encountering a hexagonally stacked pair of bilayers. Starke et al. find a best fit between experiment and theory for a 45%, 40% and 15% combination of S1, S2, and S3 stacking, and we employ the same combination (no structural parameters are given for the S3 stacking by Starke et al, but we use the same parameters for the S3 domain as the S1 domain, i.e. shifted by one bilayer).[26] The Pendry R-factor for the fit between the theory and the experiment in Fig. 5 is 0.26, indicating good agreement between experimental and theoretical intensity spectra.[21]

Returning to the oxidized buffer layer structure, dashed lines in Fig. 4 show LEED computations results carried out for a model with one additional graphene layer on top of the silicate layer. A $2\sqrt{3} \times 2\sqrt{3}$-R30±6.59° graphene commensurate structure is used for the additional graphene layer. The structure of the silicate layer is still the same as that used by Starke *et al*,[26] although we employ only the S3 stacking termination since we find that that produces the best fit with experiment (various terminations including fractional amounts of S1 and S2 have been tested, with the best fit obtained using >70% S3 termination). A graphene layer has initially a specified separation from the silicate, and the vertical coordinates of the graphene are then permitted to relax over distances of ±0.02 nm. The optimized *I(E)* curves agree well with experiment, yielding a relatively low R-factor of 0.18. This level of agreement between experimental and calculated intensities provides the main evidence for the correctness of our structural model of Fig. 1(b), with a silicate layer in the form of $Si_2O_3$ appearing between the decoupled buffer layer and the SiC substrate. The separation between the decoupled buffer layer and the oxygen atoms of the silicate layer in the results of Fig. 4 is 0.23 nm, although the R-factor is quite insensitive to this value. Our best determination of separation arises from the LEER results discussed in Section III(A), where the separation between decoupled buffer and silicate layer is found, qualitatively, to be significantly less than the 0.33 nm separation between graphene layers.

Comparing the results of Fig. 4 and Fig. 5, a noticeable difference of their *I(E)* curves occurs for the intensity of the (4/3,1/3) beam, which, relative to the (1,0) beam, is much lower for the bare oxidized surface (Fig. 4) than for the graphene-covered surface (Fig. 5). Using integrated intensities of the measured intensities, the ratio of (4/3,1/3) intensity over (1,0) intensity is only about 0.05 for the bare oxidized SiC surface, while it is about 0.2 for the graphene-covered sample. The calculated curves display similar values for these ratios. It appears that the (4/3,1/3) beam is more or less forbidden for the bare oxidized surface, i.e. due to the symmetry of the precise atomic arrangement formed in that case. With one or more additional graphene layers on top, the symmetry changes, so that the (4/3,1/3) beam is much more intense from the graphene covered surface. The approximate agreement in intensity ratio between experiment and theory is another piece of evidence for the correctness of our structural model for the decoupled graphene-like buffer layer.

## IV. Discussion

We have obtained the $\sqrt{43}$ LEED pattern on several samples prepared in 5 × 10$^{-5}$ Torr of disilane, and we have also obtained it for samples prepared in a purified neon environment, but never in vacuum. It seems the formation of this $\sqrt{43}$ interface structures requires formation conditions that are closer to equilibrium than those of vacuum, i.e. similar to the situation for graphene formation on the Si-face as argued by Tromp and Hannon.[27] Although we have not determined the exact structure of the layer between the graphene-like buffer layer and the underlying SiC crystal, as represented by the box with question marks in Fig. 1(a), it is possible that this layer contains excess Si atoms compared to a SiC bilayer. Determining the stoichiometry of this layer immediately below the buffer layer is a crucial issue for the complete structural determination of the unoxidized graphene-on-SiC surface. In any case, during the subsequent graphene formation it is expected that the graphene-like buffer layer becomes a new graphene layer and another graphene-like buffer layer forms underneath it and bonds to the substrate, in the same way as graphene growth occurs on the Si-face.[10]

After air exposure, with or without subsequent annealing in oxygen environment, the $\sqrt{43}$ pattern disappears and electron reflectivity spectra measured by LEEM change.[16] This transformation has been observed repeatedly on several samples we prepared. As already discussed above, we interpret this transformation as arising from decoupling of the graphene-like buffer layer from the underlying SiC, analogous to that which occurs for the $6\sqrt{3} \times 6\sqrt{3}$-R30° buffer layer on the Si-face.[22,23,24,25] In many cases we observe the $\sqrt{3} \times \sqrt{3}$-R30° LEED pattern to form after oxidation of the graphitized surface, but not always. By comparing the detailed air/oxygen exposure and heating conditions for all our samples, it seems that the $\sqrt{3} \times \sqrt{3}$-R30° pattern is more likely to form on samples with fewer graphene layers and higher subsequent temperature annealing. (This result is consistent with the LEER results of Fig. 3(b), in which a *decoupled* buffer is not generally observed under 2 or more layers of graphene). The formation of an ordered $Si_2O_3$ silicate layer under the graphene requires the right amount of Si and O atoms, and thick multilayer graphene may restrict the transport of O atoms through it. However, even on those samples without a $\sqrt{3} \times \sqrt{3}$-R30° pattern, LEER spectra the same as those of Fig. 3(b) are obtained (i.e. including the $G_0$ and $G_0$+G spectra of Fig. 3(b) in particular); indicating that decoupling can occur even without the formation of an ordered silicate layer.

The bonding and decoupling behavior of the buffer layer on the C-face is similar to that of the $6\sqrt{3}$ buffer layer on the Si-face. But the behavior of the two surfaces is still different in some aspects. First, the $6\sqrt{3}$ buffer layer is quite stable and can survive under many environments, whereas even with just a few days of air exposure the $\sqrt{43}$ pattern will disappear. Second, since the $6\sqrt{3}$ buffer layer acts as a template for subsequent graphene formation, graphene layers do not have rotational disorder on the Si-face. However, we still get rotational disordered graphene films on the C-face with the presence of a buffer layer, although the disorder is much less severe than for vacuum prepared samples.[14,15]

For single layer graphene on the C-face of SiC, the group of de Heer and co-workers has reported a diffraction pattern consisting of sharp graphene spots located at positions rotated by 30° relative to the principle (1,0) SiC spots.[19] We sometimes obtain a similar arrangement of graphene spots in our samples with reasonably sharp graphene spots along a 30° azimuth relative to the SiC spots[16] (these types of graphene diffraction patterns are actually quite unusual on the C-face since, as just mentioned, the graphene spots more commonly are significantly broadened due to rotational disorder[3]). However, a significant difference in the patterns from our samples compared to that of de Heer et al. is that, after oxidation, our patterns display a $\sqrt{3} \times \sqrt{3}$-R30° pattern (or a $\sqrt{43}$ pattern before oxidation) whereas the reported pattern of de Heer et al. shows no such spots.[19] Hence, it appears that no ordered silicate layer is present on their samples. Further investigation of the graphene/SiC interfaces on their material (e.g. a LEER spectrum), compared to ours, is needed to further discern possible differences in the structures.

**V. Conclusions**

By preparing graphene on the C-face of SiC in Si-rich environments, a new interface structure with $\sqrt{43} \times \sqrt{43}$-R±7.6° symmetry is found to form between graphene layers and the underlying SiC substrate. Before oxidation of the surface, the bottommost graphene layer is bonded in some way to the SiC, hence this layer forms a "buffer layer" with structure similar to that of graphene but with electronic properties that are likely quite different. After oxidation, the buffer layer decouples and becomes a graphene layer. This decoupling behavior is analogous to the decoupling of $6\sqrt{3}$ buffer layer on the Si-face. After decoupling, an ordered $Si_2O_3$ silicate layer is found to usually form between the decoupled buffer layer and the underlying SiC crystal (although the decoupling can also occur even without an *ordered* silicate layer forming, i.e. through the formation of what we believe to be a disordered oxide layer).

Acknowledgements

This work was supported by the National Science Foundation, under grant DMR-1205275.

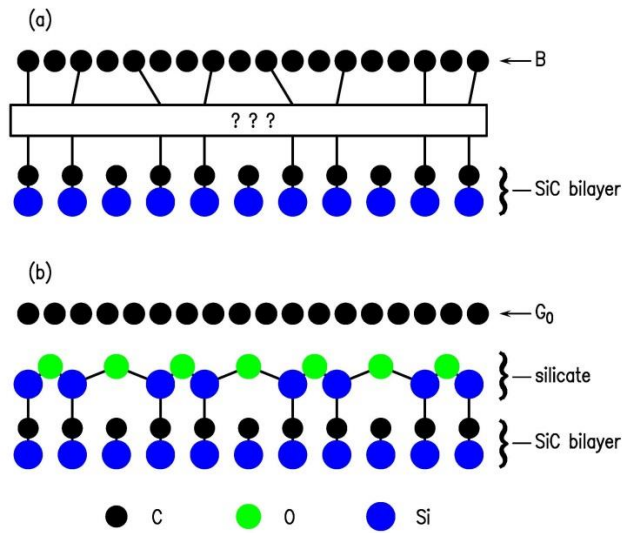

FIG. 1. Schematic view of the proposed models: (a) before oxidation, the graphene-like buffer layer (denoted as B) bonds to the underlying layer whose structure is not yet known; (b) after oxidation, the buffer layer decouples and becomes a graphene layer (denoted as $G_0$); a silicate layer with the form of $Si_2O_3$ appears between this graphene layer and the SiC substrate.

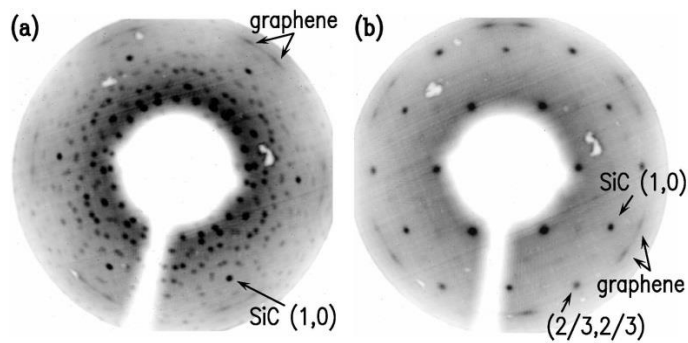

FIG. 2. LEED patterns obtained at 100 eV from 6H-SiC($000\bar{1}$) surfaces: (a) LEED pattern obtained *in situ* from a sample heated in 5 × 10$^{-5}$ Torr of disilane at 1250 °C for 5 min, showing a complex LEED pattern with graphene streaks, (b) LEED pattern obtained from an oxidized sample, showing a $\sqrt{3} \times \sqrt{3}$-R30° pattern.

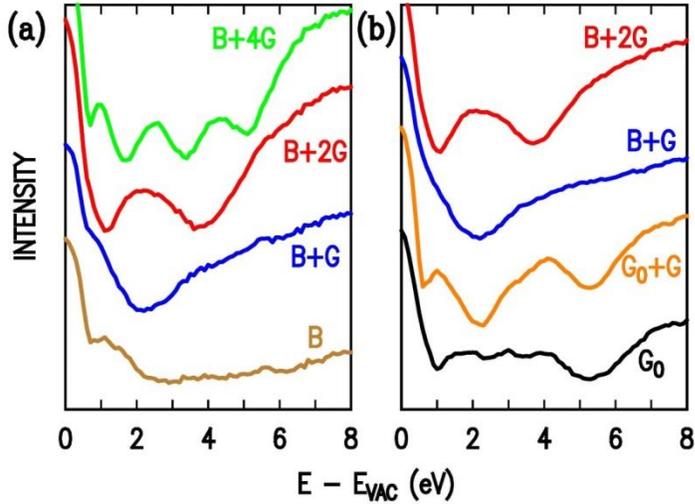

FIG. 3. LEER spectra from (a) unoxidized, and (b) oxidized surfaces of graphene on C-face SiC. Spectra are labeled according to the graphene (G) or graphene-like buffer layer (B) present on the surface, with $G_0$ denoting the buffer layer that is decoupled from the SiC and forms a regular, pristine graphene layer. The spectra have been shifted such that the vacuum level for each spectrum (as seen by the sharp increase in the reflectivity as a function of decreasing energy) is approximately aligned with zero energy.

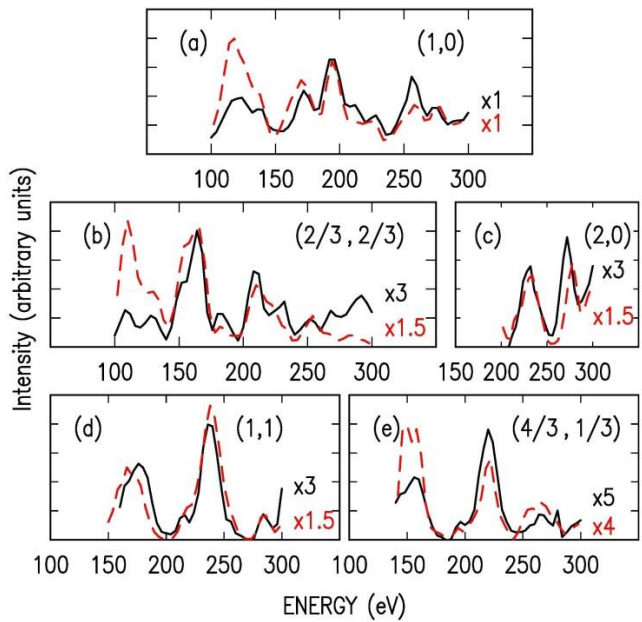

FIG. 4. Experimental LEED spot intensity spectra (solid line) obtained from the sample shown in Fig. 1(b). Dashed lines are spectra obtained from theoretical calculations. Good agreement is obtained between the experimental and theoretical spectra, as manifested by the $R$-factor of 0.18.

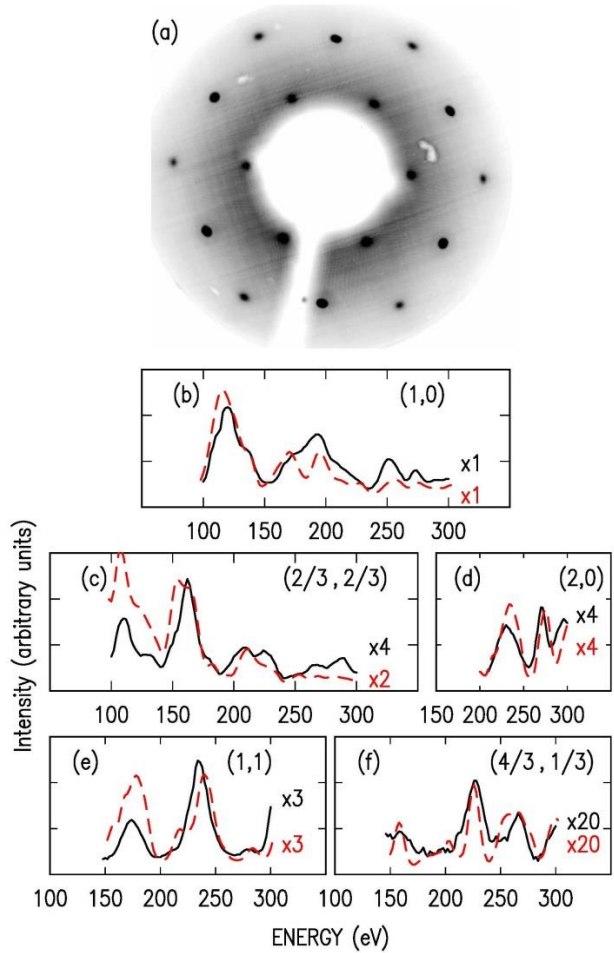

FIG. 5. (a): LEED pattern obtained from a bare oxidized 6H-SiC($000\bar{1}$) surface, prepared by annealing in argon with residual oxygen. (b)-(f): A set of experimental LEED spot intensity spectra (solid lines), together with theoretically calculated spectra (dashed lines).

References:

[1] K. S. Novoselov, A. K. Geim S. V. Morozov, D. Jiang, Y. Zhang, S. V. Dubonos, I. V. Grigorieva, and A. A. Firsov, Science **306**, 666 (2004).

[2] C. Berger, Z. Song, T. Li, X. Li, A. Y. Ogbazghi, R. Feng, Z. Dai, A. N. Marchenkov, E. H. Conrad, P. N. First, and W. A. de Heer, J. Phys. Chem. B **108**, 19912 (2004).

[3] J. Hass, W. A. de Heer, and E. H. Conrad, J. Phys.: Condens. Matter **20**, 323202 (2008).

[4] K. V. Emtsev et al., Nature Mater. **8**, 203 (2009).

[5] C. Virojanadara, M. Syväjarvi. R. Yakimova, L. I. Johansson, A. A. Zakharov, and T. Balasubramanian, Phys. Rev. B **78**, 245403 (2008).

[6] Luxmi, N. Srivastava, R. M. Feenstra, and P. J. Fisher, J. Vac. Sci. Technol. B **28**, C5C1 (2010).

[7] J. L. Tedesco et al., Appl. Phys. Lett. **96**, 222103 (2010).

[8] N. Camara, J.-R. Huntzinger, G. Rius, A. Tiberj, N. Mestres, F. Pérez-Murano, P. Godignon, and J. Camassel, Phys. Rev. B **80**, 125410 (2009).

[9] Luxmi, N. Srivastava, G. He, R. M. Feenstra, and P. J. Fisher, Phys. Rev. B **82**, 235406 (2010).

[10] K. V. Emtsev, F. Speck, Th. Seyller, L. Ley, and J. D. Riley, Phys. Rev. B. **77**, 155303 (2008).

[11] C. Riedl, U. Starke, J. Bernhardt, M. Franke, and K. Heinz, Phys. Rev. B **76**, 245406 (2007).

[12] F. Varchon, R. Feng, J. Hass, X. Li, B. N. Nguyen, C. Naud, P. Mallet, J.-Y. Veuillen, C. Berger, E. H. Conrad, and L. Magaud, Phys. Rev. Lett. **99**, 126805 (2007).

[13] F. Hiebel, P. Mallet, F. Varchon, L. Magaud, and J.-Y. Veuillen, Phys. Rev. B **78**, 153412 (2008).

[14] N. Srivastava, G. He, Luxmi, P. C. Mende, R. M. Feenstra and Y. Sun, J. Phys. D: Appl. Phys. **45,** 154001 (2010).

[15] N. Srivastava, G. He, and R. M. Feenstra, Phys. Rev. B **85,** 041404(R) (2012).

[16] G. He, N. Srivastava, and R. M. Feenstra, J. Vac. Sci. Technol. B **30**, 04E102 (2010).

[17] R. M. Feenstra, N. Srivastava, Q. Gao, M. Widom, B. Diaconescu, T. Ohta, G. L. Kellogg, J. T. Robinson, and I. V. Vlassiouk, Phys. Rev. B **87**, 041406(R) (2013).

[18] N. Srivastava, Q. Gao, M. Widom, R. M. Feenstra, S. Nie, K. F. McCarty, and I. V. Vlassiouk, Phys. Rev. B **87**, 245414 (2013).

[19] W. A. de Heer, C. Berger, M. Ruan, M. Sprinkle, X. Li, Y, Hu, B. Zhang, J. Hankinson, and E. Conrad, Proc. Natl. Acad. Sci. U.S.A. **108**, 16900 (2011).

[20] V. Blum and K. Heinz, Computer Physics Communications **134,** 392-425 (2001).

[21] J. B. Pendry, J. Phys. C **13**, 937 (1980).

[22] C. Riedl, C. Coletti, T. Iwasaki, A. A. Zakharov, and U. Starke, Phys. Rev. Lett. **103**, 246804 (2009).

[23] S. Oida, F. R. McFeely, J. B. Hannon, R. M. Tromp, M. Copel, Z. Chen, Y. Sun, D. B. Farmer, and J. Yurkas, Phys. Rev. B **82**, 041411(R) (2010).

[24] C. Virojanadara, S. Watcharinyanon, A. A. Zakharov, and L. I. Johansson, Phys. Rev. B **82**, 205402 (2010).

[25] K. V. Emtsev, A. A. Zakharov, C. Coletti, S. Forti, and U. Starke, Phys. Rev. B **84**, 125423 (2011).

[26] U. Starke, J. Schardt, J. Bernhardt, and K. Heinz, J. Vac. Sci. Technol. A **17**, 1688 (1999).

[27] R. M. Tromp and J. B. Hannon, Phys. Rev. Lett. **102**, 106104 (2009).